\documentclass[namedreferences]{kapproc}\usepackage{latexsym}
\usepackage{epsfig}
\newfont\jafont{wncyr10} % 10 pt
\newcommand{\rhdya}{\mbox{\jafont YA}}
\newcommand{\rhdmulti}[1]{\mbox{\underline{\bf #1}}}
\newcommand{\rhddeltp}[2]{\Delta^{#1}_{\frac{#2}{2}}}
\newcommand{\rhddeltm}[2]{\Delta^{#1}_{-\frac{#2}{2}}}
\newcommand{\rhdnuklp}[1]{\mbox{N}^{#1}_{\frac{1}{2}}}
\newcommand{\rhdnuklm}[1]{\mbox{N}^{#1}_{-\frac{1}{2}}}
\newcommand{\rhdne}{\sqrt{\frac{1}{3}}}
\newcommand{\rhdnz}{\sqrt{\frac{2}{3}}}
\runningtitle{Relativistic SU(4) and Quaternions}
\runningauthor{Rolf Dahm}

\begin{opening}
\title{Relativistic SU(4) and Quaternions}
\subtitle{International Conference on the Theory of the Electron,\\
ICTE-1995, Cuautitl\'an, M\'exico}
\author{Rolf Dahm}
\institute{Institute of Nuclear Physics\\
University Mainz\\
D-55099 Mainz, Germany\\
dahm@vkpmzj.kph.uni-mainz.de}
\end{opening}

\begin{document}
%\begin{center}Received 1 November 1995 [deadline]\end{center}
\begin{center}Received 1 November 1995 \end{center}

\hyphenation{qua-ter-ni-on}

\begin{abstract}
A classification of hadrons and their interactions at low energies according 
to SU(4) allows to identify combinations of the fifteen mesons $\pi$, $\omega$ 
and $\rho$ within the spin-isospin decomposition of the regular representation 
\rhdmulti{15}. Chirally symmetric SU(2)$\times$SU(2) hadron interactions are 
then associated with transformations of a subgroup of SU(4). Nucleon and Delta 
resonance states are represented by a symmetric third rank tensor \rhdmulti{20} 
whose spin-isospin decomposition leads to $4\oplus 16$ `tower states' also 
known from the large-N$_c$ limit of QCD. Towards a relativistic hadron theory, 
we consider possible generalizations of the stereographic projection 
{\bf S}$^{2}$ $\to$ {\bf C} and the related complex spinorial calculus {\it on 
the basis of the division algebras with unit element}. Such a geometrical 
framework leads directly to transformations in a quaternionic projective 
`plane' and the related symmetry group SL(2,{\bf H}). In exploiting the Lie 
algebra isomorphism sl(2,{\bf H}) $\cong$ su$*$(4) $\cong$ so(5,1), we focus 
on the Lie algebra su$*$(4) to construct quaternionic Dirac-like spinors, the 
associated Clifford algebra and the relation to SU(4) by Weyl's unitary trick. 
The algebra so(5,1) contains the de Sitter-algebra so(4,1) which can be 
contracted to the algebra of the Poincar\'e group.
\end{abstract}

\noindent{\footnotesize\bf PACS numbers:} {\footnotesize }\\
\noindent{\footnotesize\bf 1991 Mathematics Subject Classification:}
{\footnotesize }\\

\tableofcontents

\section{Introduction}
\label{sect1}

Nowadays, QCD is assumed to be the theory of hadronic interactions at high 
energies, however, there is no complete and consistent theory for hadrons 
at low and intermediate energies yet. Instead of trying to `reduce' QCD 
towards a low energy theory of hadrons or dealing with `QCD inspired' 
effective hadron models, we shall start with symmetry properties and quantum 
numbers known from hadron dynamics at low energies, especially from the 
$\pi N\Delta$ system. On this basis, we can generalize the known symmetry 
properties of hadrons towards a relativistic quantum field theory. 

\section{Classification of hadrons}
\label{sect2}

The simplest hadronic classification scheme is based on the group SU(2) of 
isospin symmetry which is realized in the Wigner-Weyl mode. However, when 
calculating pion-nucleon scattering processes \cite{adldash}, it became 
apparent that this `static' classification scheme is not sufficient to yield 
a complete description of pion dynamics but that it has to be extended by the 
more sophisticated concept of a spontaneously broken symmetry.

\subsection{Chiral Dynamics}
% kurz die Aussagen der Chiralen Dynamik

This more sophisticated approach to hadron dynamics is mainly based on an 
underlying `chiral symmetry', described by the group SU(2)$\times$SU(2). 
Spontaneous breakdown of this symmetry can be treated in terms of projections 
with respect to the diagonal subgroup SU$_{V}$(2) which has been identified 
with the isospin symmetry group \cite{adldash}, \cite{weinberg1}. 
Mathematically, this approach towards SU$_{V}$(2) isospin quantum numbers is 
well established in the framework of the coset decomposition 
SU(2)$\times$SU(2)/SU$_{V}$(2) \cite{wesszum1},\cite{wesszum2}. However, 
physically these so called `nonlinear sigma models' \index{nonlinear sigma 
models} raise a lot of serious problems. The general inherent problem results 
from the identification of physical particles with representations of the 
subgroup SU$_{V}$(2) so that the action of `chiral group transformations' on 
irreducible (static) isospin representations of the field algebra generates 
{\it inequivalent} representations \cite{joosweim} and thus changes the 
properties and the quantum numbers of the particle. Furthermore, not all 
automorphisms can be realized by {\it unitary} operators on linear 
representation spaces of the subgroup SU$_{V}$(2) \cite{haka}, \cite{fabpic}. 
Thus, due to the identification of SU$_{V}$(2) representations with physical 
particles, this concept leads necessarily to highly nonlinear models with 
enormous mathematical complications and to a loss of renormalizability. A 
typical SU(2)$\times$SU(2) nonlinear sigma model \cite{leutwyler1} describes 
the mesons by means of 
\begin{equation}
\label{eq:start}
U\,=\,\exp\left(-i\gamma_{5}\vec{\tau}\cdot\vec{\varphi}/f_{\pi}\right)\,.
\end{equation}
The related `quantum field theory' is based on an effective (nonlinear) 
Lagrangian in terms of `fields' $U$ and (covariant) derivatives $\nabla^{\mu}U$ 
where the `field' $U$ is usually expanded into a power series (for details, 
see \cite{leutwyler1} and references therein). Furthermore, nucleon resonances 
like the Delta isobar excitation are usually not considered in these models 
although it is known that both nucleon and Delta states are needed to saturate 
the Adler-Weisberger sum rule \cite{oehme}, \cite{kiri}.

\subsection{Dynamic classification of the nucleon}
\label{dynamclass}
% Sudarshan & Nukleon\neq Dirac-Nukleon
In almost all effective classification schemes and models, the nucleon 
transforms relativistically according to the same Dirac representation 
as the electron. However, with respect to its interactions and its dynamic 
properties this description of the nucleon doesn't work. Therefore, one has to 
include electromagnetic corrections from the very beginning like the Pauli 
(spin) term to correct the nucleon's magnetic moments. Furthermore, the idea 
of a `composite particle' is necessary to explain obvious deviations from the 
simple description using a fundamental Dirac spinor and to justify the use of 
(effective) formfactors in the description of photon interactions with the 
nucleon.

To avoid these phenomenological corrections to the (covariant) Dirac 
representation and the related problems, we want to revive and extend an 
investigation concerning the properties of dynamic nucleons in the framework 
of a Goldstone realization \index{Goldstone realization} of pions 
\cite{sudarsh}. Trying to explain renormalization effects of the value of the 
axial coupling constant $g_{A}$, Sudarshan parametrized the dynamic nucleon 
according to
\begin{equation}
\label{eq:sudarshansatz}
\left|N\right>_{\mbox{\scriptsize dyn}}\,=\,
a\left|N\right>_{\mbox{\scriptsize stat}}\,+\,
\sqrt{1-a^{2}}\,\int\mathrm{d}\omega\,\pi(\omega)
\left|N(\omega)\right>_{\mbox{\scriptsize stat}}\,.
\end{equation}
The second part describes dynamic deviations of the static nucleon 
classification in terms of an integral over the nucleon/Goldstone pion 
continuum. The requirement $g_{A}=1.26$ fixes the value of $a^{2}$ in the 
parametrization (\ref{eq:sudarshansatz}) to $a^{2}\approx 2/3$. 

Technically, this approach can be equally well understood in a quasiparticle 
picture \cite{daki1} if we reinterprete the quantum numbers of the coupled 
$\vec{\pi}N$-system in eq.~(\ref{eq:sudarshansatz}) in terms of the (static) 
Delta resonance \index{Delta resonance}. Appropriately, pure SU$_{V}$(2) 
fermion representations can be used as basic constituents of `mixed' dynamic 
fermion representations which naturally occur in the quasiparticle picture. 
In order to fix the (chiral) dynamic eigenvectors of the $N\Delta$-system, 
the saturation of the Adler-Weisberger sum rule \cite{oehme}, \cite{kiri} with 
nucleon and Delta states suggests to diagonalize the nucleon/Delta axial 
charge matrix,
\begin{equation}
Q_{5}\,=\,\left(
\begin{array}{cc}
<\rhdnuklp{+}\,|A_{3}|\,\rhdnuklp{+}>\raisebox{-3mm}{\rule{0mm}{5mm}} & 
<\rhddeltp{+}{1}\,|A_{3}|\,\rhdnuklp{+}>\\
<\rhdnuklp{+}\,|A_{3}\,|\,\rhddeltp{+}{1}>\raisebox{-2mm}{\rule{0mm}{4mm}} & 
<\rhddeltp{+}{1}\,|A_{3}|\,\rhddeltp{+}{1}>
\end{array}
\right)\,.
\end{equation}
This diagonalization leads to the same structure of dynamic states as in 
Sudarshan's picture \cite{daki1}, and the states (\ref{eq:sudarshansatz}) are 
included in a more general set of dynamic fermion representations (`Chirons'). 

Due to the quasiparticle picture, it is obvious that starting with irreducible 
SU$_{V}$(2) fermion representations the dynamics of chiral transformations 
will mix these states so that isospin will no longer be conserved. Furthermore, 
this ansatz suggests with respect to spin degrees of freedom that dynamic 
`nucleons' should {\it not} be associated with a fundamental Dirac spinor but 
with a higher multiplet structure which yields nucleonic 1/2 components as 
well as 3/2 (Delta) components. Such a description avoids the inconcistency 
that operators of the chiral algebra generating the compact symmetry group 
SU(2)$\times$SU(2) need to connect {\it different} irreducible fermion 
representations $N$ and $\Delta$ of the isospin as well as the chiral group.

\subsection{SU(4) classification scheme}
\label{su4class}
% Verallgemeinerung auf SU(4)

The ideas summarized in the last sections can be realized in a group theoretic 
framework if we find linear representations to classify quantum numbers of 
fermions and mesons in the physical spectrum which allow to represent all 
actions of the complete algebra of Chiral Dynamics\index{Chiral Dynamics}. As 
mentioned in section~\ref{dynamclass}, the saturation of the Adler-Weisberger 
sum rule leads to the qualitative suggestion to describe SU$_{V}$(2) nucleon 
and Delta representations in the {\it same representation} so that the `chiral' 
algebra can `connect' these states when acting on a fermion multiplet. 
Sudarshan's work resp. the quasiparticle picture allows for quantitative 
statements about the `mixture' of nucleon and Delta fermions by fixing the 
parameter $a$ introduced in (\ref{eq:sudarshansatz}).

An ansatz on the basis of the Lie group SU(4) \index{SU(4)} realizes these 
suggestions if we interpret SU(4) in terms of spin-flavour transformations and 
respect the SU(4) reduction according to the chain SU(4) $\supset$ USp(4) 
($\cong$ Sp(2)) $\supset$ SU(2)$\times$SU(2) $\supset$ SU(2). In this 
framework, dynamic nucleon and Delta states of Chiral Dynamics are 
represented by the third rank symmetric spinor representation \rhdmulti{20} of 
spin-flavour SU(4), $\Psi^{\alpha\beta\gamma}$, $1\leq\alpha,\beta,\gamma\leq 
4$, which reduces with respect to spin-isospin quantum numbers \cite{daki1} 
according to
\begin{equation}
\label{eq:fermequat}
\begin{array}{lclclcl}
\Psi^{111} & = & \rhddeltp{++}{3}, 
& \quad &
\Psi^{114} & = & \rhdnz\,\rhdnuklp{+}+\rhdne\,\rhddeltp{+}{1},
\\

\Psi^{113} & = & \rhddeltp{++}{1}, 
& \quad &
\Psi^{144} & = & \rhdnz\,\rhdnuklm{0}+\rhdne\,\rhddeltm{0}{1},
\\

\Psi^{133} & = & \rhddeltm{++}{1}\,, 
& \quad & 
\Psi^{134} & = & \rhdne\,\rhdnuklm{+}+\rhdnz\,\rhddeltm{+}{1},
\\

\Psi^{333} & = & \rhddeltm{++}{3}, 
& \quad &
\Psi^{124} & = & \rhdne\,\rhdnuklp{0}+\rhdnz\,\rhddeltp{0}{1},
\\

\Psi^{112} & = & \rhddeltp{+}{3}, 
& \quad &
\Psi^{123} & = & -\rhdne\,\rhdnuklp{+}+\rhdnz\,\rhddeltp{+}{1},
\\

\Psi^{334} & = & \rhddeltm{+}{3},
& \quad & 
\Psi^{234} & = & -\rhdne\,\rhdnuklm{0}+\rhdnz\,\rhddeltm{0}{1},
\\

\Psi^{221} & = & \rhddeltp{0}{3}, 
& \quad & 
\Psi^{332} & = & -\rhdnz\,\rhdnuklm{+}+\rhdne\,\rhddeltm{+}{1},
\\

\Psi^{443} & = & \rhddeltm{0}{3}, 
& \quad &
\Psi^{223} & = & -\rhdnz\,\rhdnuklp{0}+\rhdne\,\rhddeltp{0}{1},
\\

\Psi^{222} & = & \rhddeltp{-}{3}, 
& \quad &
\Psi^{224} & = & \rhddeltp{-}{1},
\\

\Psi^{244} & = & \rhddeltm{-}{1},
& \quad & 
\Psi^{444} & = & \rhddeltm{-}{3}\,.
\end{array}
\end{equation}
The upper index denotes the charge and the lower one the spin projection 
of the state. In symbolic notation this reduction reads as 
\begin{equation}
\rhdmulti{20}\quad\longrightarrow\quad\left({1\over 2},{1\over 2}\right)\,\oplus
\,\left({3\over 2},{3\over 2}\right)\,.
\end{equation}
which yields exactly the tower states also obtained in the large-$N_{c}$ limit 
of QCD. 

The {\it massive} mesons $\vec{\pi}$, $\omega$ and $\vec{\rho}$ 
are identified within the regular representation \rhdmulti{15} of SU(4), 
$M^{\alpha\dot{\beta}}$, which can be decomposed according to 
\begin{equation}
\rhdmulti{15}\quad\longrightarrow\quad
\left(0,1\right)\,\oplus\,\left(1,0\right)\,\oplus\,\left(1,1\right)
\end{equation}
with respect to spin-isospin degrees of freedom. The explicit representation 
of the meson vector is given by
\begin{equation}
\label{eq:mesonequat}
\begin{array}{rclcrcl}
M^{1\dot{1}} & = & {1\over 2}(\pi^{0}+\omega^{0}+\rho^{0}_{0}),
& \quad &
M^{1\dot{2}} & = & \sqrt{{1\over 2}}(\pi^{+}+\rho^{+}_{0}),
\\ 
M^{1\dot{3}} & = & \sqrt{{1\over 2}}(\omega_{+}+\rho^{0}_{+}),
& \quad &
M^{1\dot{4}} & = & \rho^{+}_{+},
\\ 

M^{2\dot{1}} & = & \sqrt{{1\over 2}}(\pi^{-}+\rho^{-}_{0}),
& \quad &
M^{2\dot{2}} & = & {1\over 2}(-\pi^{0}+\omega_{0}-\rho^{0}_{0}),
\\ 
M^{2\dot{3}} & = & \rho^{-}_{+},
& \quad &
M^{2\dot{4}} & = & \sqrt{{1\over 2}}(\omega_{+}-\rho^{0}_{+}),
\\

M^{3\dot{1}} & = & \sqrt{{1\over 2}} (\omega_{-}+\rho^{0}_{-}),
& \quad &
M^{3\dot{2}} & = & \rho^{+}_{-},
\\ 
M^{3\dot{3}} & = & {1\over 2}(\pi^{0}-\omega_{0}-\rho^{0}_{0}),
& \quad &
M^{3\dot{4}} & = & \sqrt{{1\over 2}}(\pi^{+}-\rho^{+}_{0}),
\\ 

M^{4\dot{1}} & = & \rho^{-}_{-},
& \quad &
M^{4\dot{2}} & = & \sqrt{{1\over 2}}(\omega_{-}-\rho^{0}_{-}),
\\ 
M^{4\dot{3}} & = & \sqrt{{1\over 2}}(\pi^{-}-\rho^{-}_{0}),
& \quad &
M^{4\dot{4}} & = & {1\over 2}(-\pi^{0}-\omega_{0}+\rho^{0}_{0}),
\end{array}
\end{equation}
All these states have good spin and isospin projections $S_{3}$ and $T_{3}$
(electromagnetic charge), however, $S^{2}$ and $T^{2}$ are no Casimir 
operators of the rank 3 group SU(4), and neither total spin nor isospin are 
conserved in an SU(4) symmetric theory. It is wellknown that spin is not 
conserved in relativistic dynamics, i.e. $S^{2}$ is not an appropriate Casimir 
operator with respect to a relativistic particle classification scheme. 
Nonconservation of isospin is already known from Chiral Dynamics where axial 
transformations `connect' irreducible mesonic and fermionic isospin 
representations. Besides the axial transformations acting on the fermion space 
spanned by the $N\Delta$-system, it is wellknown that in the linear sigma model 
\cite{gmlevy} the axial generators $X_{j}$ connect an isospin singlet state 
$\sigma$ ($T^{2}\sigma=0$) with isospin triplet states $\vec{\pi}$ 
($T^{2}\vec{\pi}=t(t+1)\vec{\pi}=2\vec{\pi}$) states according to the 
commutation relations
\begin{equation}
\left[X_{j},\sigma\right]\,=\,i\,\pi_{j}\,,\quad
\left[X_{j},\pi_{k}\right]\,=\,-i\,\delta_{jk}\sigma.
\end{equation}
The generators $T_{j}$ of the isospin subgroup SU$_{V}$(2) do not interchange 
the meson representations,
\begin{equation}
\left[T_{j},\sigma\right]\,=\,0\,,\quad
\left[T_{j},\pi_{k}\right]\,=\,i\,\epsilon_{jkl}\pi_{l}\,.
\end{equation}

In a SU(4) hadron theory, the nonconservation of isospin has 
no influence on the definition of electromagnetic charges due to a well 
defined projection $T_{3}$, however, isospin symmetry is slightly broken 
already by pure hadronic interactions. For example, the coupling of charged 
and neutral pions to the nucleon differs by $\approx$ 10\% \cite{daki1}, 
\cite{dahmdiss} if we identify physical hadronic states (in analogy to Chiral 
Dynamics) by the quantum numbers of the isospin reduction (\ref{eq:mesonequat}).

Using (\ref{eq:fermequat}) and (\ref{eq:mesonequat}), a linear meson-fermion 
vertex \index{SU(4) meson-fermion vertex} describing hadronic interactions at 
low energies may be constructed according to the standard rules of SU(4) tensor 
algebra as \cite{daki1}, \cite{dahmdiss}
\begin{equation}
{\mathcal L}^{\mbox{\scriptsize int}}\,=\,
G\,J^{\dot{\alpha}\beta}\,M^{\alpha\dot{\beta}}\,=\,
G\,\Psi^{\dot{\alpha}\dot{\gamma}\dot{\delta}}\Psi^{\beta\gamma\delta}\,
M^{\alpha\dot{\beta}}\,.
\end{equation}
The algebra su(2)$\oplus$su(2) of Chiral Dynamics can be identified as a 
subalgebra of the Lie algebra su(4) and acts on the irreducible su(4) 
representations $\Psi^{\alpha\beta\gamma}$ and $M^{\alpha\dot{\beta}}$ 
\cite{dahmdiss}. Some results of the SU(4) coupling scheme and their comparison 
with experiments are given in \cite{daki1}, \cite{dahmdiss}. However, as in 
the case of Chiral Dynamics it should be noted that the compact symmetry group 
SU(4) yields a good description of hadronic properties only at very low 
energies. With respect to a relativistic quantum field theory, it is at least 
necessary to find an appropriate noncompact symmetry group.

\subsection{Towards a relativistic hadron theory}
% Wigner & large-Nc

Before looking for such a noncompact symmetry group, it is noteworthy to 
mention some results obtained long ago from a completely different dynamic 
physical system, namely the classification scheme of ground states of nuclei.

In this context, Wigner \cite{wigner1} has already shown in 1937 that the Lie 
group SU(4) allows a reasonable classification of ground states. During the 
early 60ies, it has been shown on the basis of more complete experimental data 
that SU(4) indeed yields a good description of ground states of nuclei in the 
range of A=1 to A=140 \cite{franz1}, \cite{pais1}. However, as soon as energy 
raises and dynamic effects become more important, SU(4) symmetry becomes worse.

This `SU(4) behaviour' of the dynamically completely different system of 
nuclear ground states at low energies suggests to look for noncompact groups 
related to SU(4) by Weyl's unitary trick so that {\it at very low energies} 
the various experiments cannot distinguish between SU(4) transformations and 
its related noncompact `counterpart(s)'. We are thus led to the group SU$*$(4) 
\index{SU$*$(4)} which results from embedding quaternions \index{quaternions} 
into complex vector spaces 
\cite{helgason1} by 
\begin{equation}
\label{eq:pauliquats}
q_{0}=1_{2\times 2}\,,\quad q_{j}=-i\sigma_{j}\,,\quad 1\leq j\leq 3\,,
\end{equation}
where $\sigma_{j}$ denote the Pauli matrices.

However, besides these phenomenological considerations, based on the close 
relation of SU(4) and SU$*$(4) at low energies, there exists a straightforward 
mathematical approach. This direct approach towards a relativistic hadron 
theory is mainly inspired by eq.~(\ref{eq:start}) which can be rewritten as
\begin{equation}
\label{eq:zentral}
U\,=\,\exp\left(\vec{q}\cdot\vec{\varphi}\right)\,=\,
\cos\varphi+\vec{q}\cdot\hat{\varphi}\sin\varphi\,,\quad
\varphi\,=\,|\vec{\varphi}|\,,\quad
\hat{\varphi}\,=\,\frac{\vec{\varphi}}{\varphi}
\end{equation}
using the $2\times 2$ complex matrix representations of quaternions as given 
by eq.~(\ref{eq:pauliquats}) and their multiplication law 
$q_{j}q_{k}=-\delta_{jk}q_{0}+\epsilon_{jkl}q_{l}$. For the sake of simplicity, 
we omitted in eq.~(\ref{eq:zentral}) the rescaling $f_{\pi}$ of the parameters 
$\vec{\varphi}$ as a redefinition of the `fields' $\varphi_{j}$. Furthermore, 
with respect to the discussion of bosonic properties, we omitted the Dirac 
matrix $\gamma_{5}$ because of $\gamma_{5}^{2}=1$ and because $\gamma_{5}$ 
acts only on fermion spinors by exchanging upper and lower components. This 
exchange of the spinor components will be absorbed by the geometrical theory 
of the more general quaternionic transformations in section~\ref{quaternproj} 
so that the correspondence of $\gamma_{5}$ with the `fields' $\varphi_{j}$ is 
related to reflections of quaternions at the quaternionic unit circle, 
$q\to q^{-1}$.

With the identity given in eq.~(\ref{eq:zentral}), nonlinear realizations $U$ 
like eq.~(\ref{eq:start}) which transform according to the representation 
$({1\over 2},{1\over 2})$ of SU(2)$\times$SU(2) are nothing else but real 
quaternions normalized to unity, 
\begin{equation}
\label{eq:quaternionnorm}
||U||^{2}\,=\,\frac{1}{2}\,\mbox{\bf Tr}\left(U^{+}U\right)\,=\,
\cos^{2}\varphi\,+\,\sin^{2}\varphi\,=\,1\,.
\end{equation}
In analogy to the `polar' and the `linear' (cartesian) representation of 
complex numbers normalized to unity, $z=\exp(i\alpha)=x+iy$, $x=\cos\alpha$, 
$y=\sin\alpha$, the two {\it identical} representation schemes of the unit 
quaternion $U$ in eq.~(\ref{eq:zentral}) can be denoted as `polar' 
(`nonlinear') and `linear' (cartesian) representations. The relevance of 
SU(2)$\times$SU(2) and SO(4) transformations investigated in the framework of 
nonlinear \cite{weinberg1}, \cite{wesszum1}, \cite{wesszum2} resp. linear 
sigma models \cite{gmlevy} then becomes obvious (section~\ref{chirdyn}; 
\cite{daki2}) due to the properties of the four dimensional projection plane 
and the isomorphism U(1,{\bf H}) $\cong$ SU(2,{\bf C}). Thus, investigations 
of SU(2)$\times$SU(2) nonlinear sigma models suggest strongly to identify 
`chiral' transformations of hadron fields in the more general framework of 
quaternionic transformations. 

After a short review of the wellknown complex case which leads to SU(2) spin 
in quantum mechanics and SL(2,{\bf C}) `relativistic' spinor theory, the 
necessary generalization is given for the quaternionic case. 

\section{Noncompact groups and spinors}
\label{sect3}

\subsection{$S^2$ $\to$ {\bf C}, the complex case}
% Stereographische Projektion

The treatment of spinors in quantum mechanics is closely related to the 
stereographic projection \index{stereographic projection} $S^2$ $\to$ {\bf C} 
\cite{gms}, \cite{penrind}. Each point of the sphere $S^2$ can be described by 
two real parameters (angles). If we use the geometry given in \cite{penrind} 
where the equator of the sphere lies in a plane parametrized by two real 
parameters (cartesian coordinates), the stereographic projection associates a 
point $P$ on the sphere with the two real coordinates of a point $P'$ of the 
equatorial plane which denotes the intersection of a line passing through the 
north pole $N$ and the point $P\in S^{2}$. If the point $P$ on the sphere moves 
on a continuous curve through the north pole $N$, the projection $P'$ has to 
move through {\it one} infinite point in the plane. This closure including 
{\it one} infinite point is possible by a relative complexification $i$ of the 
two real planar cartesian coordinates, and transformations of $P'$ can be 
investigated by appropriate transformations of complex numbers in the complex 
`plane'\footnote{We'll denote the projective one dimensional complex and 
quaternionic `lines' by `planes' although the nomenclature `plane' is justified 
only in the complex case with respect to the involved two real parameters.} 
{\bf C}. In this context it is noteworthy, that the projection $S^{2}$ $\to$ 
{\bf C} {\it automatically} leads to a commutative theory independent of the 
character and the further interpretation of the hypercomplex unit $i$ since we 
use only {\it one} hypercomplex unit as a relative complexification of the two 
real planar coordinates. Therefore, this `minimal' commutative approach to a 
projective theory as used in quantum mechanics doesn't justify a priori the 
assumption of $i$ to be commutative when we embed this complex projective 
theory into a higher hypercomplex number system. 

The projective transformations can be described by M\"obius transformations 
\index{M\"obius transformations} in the complex plane,
\begin{equation}
\label{eq:moebius}
f(z)\,\to\,f'(z)\,=\,\frac{\alpha z+\beta}{\gamma z+\delta}\,,\quad
\alpha,\beta,\gamma,\delta\in\mbox{\bf C}\,,
\end{equation}
so that curves on $S^{2}$ are mapped onto curves in {\bf C}. In {\bf C} one 
benefits from complex analysis and well defined contour integrals which are 
related to finite paths on $S^{2}$, a nice feature which is for example used 
when applying dispersion relations in physics. The possibility to `close' these 
paths by adding {\it one} infinite point is intimately related to the fact 
that eq.~(\ref{eq:moebius}) has exactly {\it one} singularity, i.e. that the 
equation 
\begin{equation}
\label{eq:huepf}
\gamma z+\delta\,=\,0\,,\quad \gamma,\delta\in\mbox{\bf C}\,,
\end{equation}
derived from the denominator of (\ref{eq:moebius}) has an {\it unique} 
solution. However, the geometry described by M\"obius transformations 
(\ref{eq:moebius}) may be equally well treated in terms of matrix algebras by 
the identification
\begin{equation}
\label{eq:gruppe}
f(z)\,=\,\frac{\alpha z+\beta}{\gamma z+\delta}\quad\longleftrightarrow\quad
A\,=\,\left(\begin{array}{cc}\alpha&\beta\\\gamma&\delta\end{array}\right)
\,,\quad\alpha,\beta,\gamma,\delta\in\mbox{\bf C}\,,
\end{equation}
which is formally motivated by the introduction of homogeneous coordinates 
$z\to z_{1}/z_{2}$ in eq.~(\ref{eq:moebius}). Thus, two identical formalisms 
are available to treat the relevant transformations:
\begin{itemize}
\item The use of coordinates $z$ to describe points in the complex `plane' 
allows to define an involution $\overline{z}$ (complex conjugation). This 
involution in {\bf C}, when restricted to the unit circle $|z|=1$, is 
equivalent to the replacement $z\to z^{-1}$. Transformations are described 
by M\"obius transformations (\ref{eq:moebius}) and the appropriate complex 
analysis. Furthermore, M\"obius transformations constitute a group with 
respect to composition.
\item Equivalently, one may define two dimensional complex spinors 
\[\psi\,=\,\left(\begin{array}{c}z_{1}\\z_{2}\end{array}\right)\] when using 
homogeneous coordinates $z_{1}$ and $z_{2}$. The appropriate spinorial 
transformations can be identified according to eq.~(\ref{eq:gruppe}) with 
matrices $A$, $A$ $\in$ {\bf C}$_{2\times 2}$, so that the composition of 
M\"obius transformations (\ref{eq:moebius}) is equivalent to simple matrix 
multiplication. The related matrix groups in the complex case $S^2$ $\to$ 
{\bf C} are the compact group SU(2) with respect to rotations of the sphere 
and the noncompact group SL(2,{\bf C}) with respect to general (noneuclidean) 
transformations. Restricting rotations of the sphere $S^{2}$ to rotations 
with the fixed cartesian axis $z$ (choosing the `quantization axis' $\sim$ 
$\hat{z}$) and complexifying $y$ relative to $x$, this geometry leads to the 
groups SO(2) and U(1,{\bf C}). On the representation space of square-integrable 
functions this spontaneous symmetry breaking leads to the decomposition of 
spherical harmonics $Y_{lm}(\theta,\varphi)$ in terms of (nonlinear) Legendre 
polynomials and exponentials as representations of U(1,{\bf C}).
\end{itemize}
If we want to generalize this concept on the basis of a projective `plane' 
\index{projective `plane'} and a noneuclidean geometry, it is {\it not} 
straightforward to generalize nothing but the matrix formalism on the basis of 
the related group theory to SU($n$) or SL($n$,{\bf C}), $n>2$. However, with 
respect to eq.~(\ref{eq:huepf}) derived from the denominator of the M\"obius 
transformation, it is obvious that for the `numbers' $\gamma$, $z$ and 
$\delta$ multiplication as well as addition has to be defined. To avoid 
problems with zero divisors of the necessary algebra and in order to add only 
{\it one} infinite point, the simplest possible generalization is a 
generalization on the basis of {\it division algebras}\index{division algebra}. 
This leads directly to the use of quaternions in eq.~(\ref{eq:moebius}) and 
the related projection $S^4$ $\to$ {\bf H}.

\subsection{$S^4$ $\to$ {\bf H}, the quaternionic case}
\label{quaternproj}
% Verallgemeinerung auf die Quaternionen

Projections from the sphere $S^{4}$ can be understood on the same geometrical 
footing as in the projection $S^{2}\to\mbox{\bf R}^{2}$ and the additional 
complexification to {\bf C}. In the case $S^{4}\to\mbox{\bf H}$, however, it 
is the symmetry group SO(5) which acts transitively on $S^{4}$. Furthermore, 
fixing the projection point at the intersection of $S^{4}$ with the fifth 
cartesian (real) axis, the remaining compact symmetry group in the projection 
plane with respect to rotations restricted to the fifth axis is SO(4). The 
corresponding spinors can be defined by a complexification of all four real 
variables {\it relative} to each other which leads now to 
{\it non\/}commutative hypercomplex units constituting the division algebra of 
quaternions. The spinors related to restricted rotations, i.e. to circles in 
the projection plane and to the orthogonal symmetry group SO(4), correspond 
to U(1,{\bf H})$\times$U(1,{\bf H}) transformations of the quaternions (see 
section~\ref{chirdyn}). However, since U(1,{\bf H}) is isomorphic to the group 
SU(2,{\bf C}), the spinor representations of the SO(4) linear sigma model can 
also be defined in terms of the complex covering group SU(2)$\times$SU(2). A 
similar geometry holds in the complex case (in quantum mechanics), where SO(3) 
acts transitively on $S^{2}$ and appropriate spinors can be defined using the 
covering group SU(2) or the noncompact group SL(2,{\bf C}) whereas in the case 
of restricted rotations of $S^{2}$ with a fixed $z$-axis the compact groups 
SO(2) resp. U(1,{\bf C}) describe the symmetry transformations in the complex 
plane.

Now, if we generalize eq.~({\ref{eq:moebius}}) to the division algebra of 
quaternions, a fraction of quaternions has to be defined carefully due to 
their noncommutativity. A suitable definition can be introduced by
\begin{equation}
\label{eq:quatmoeb}
f(q)\,=\,\frac{a q+b}{c q+d}\,:=\,
\begin{array}{c}
\multicolumn{1}{c|}{a q+b}\\\hline\multicolumn{1}{|c}{c q+d}
\end{array}
\,\equiv\,\left(a q+b\right)\,\left(c q+d\right)^{-1}
\end{equation}
with $a,b,c,d,q$ $\in$ {\bf H}. Using this definition, it is possible to handle 
the relevant quaternionic M\"obius transformations in analogy to the complex 
case in two equivalent formalisms:\index{quaternionic spinors}
\begin{itemize}
\item The use of coordinates $q$ to describe points in the quaternionic 
projective `plane' allows to define an involution $\overline{q}$ (quaternionic 
conjugation). Quaternionic transformations in the plane are described by 
generalized M\"obius transformations (\ref{eq:quatmoeb}) and an appropriate
quaternionic analysis. The transformations (\ref{eq:quatmoeb}) constitute a 
group with respect to composition, too. Like in the case of complex numbers, 
the equation $c q+d=0$ derived from the denominator in eq.~(\ref{eq:quatmoeb}) 
has an {\it unique} solution $q$ so that it is only necessary to take care 
of {\it one} singularity in eq.~(\ref{eq:quatmoeb}).
\item As a second description, one may define two dimensional quaternionic 
spinors 
\begin{equation}
\label{eq:fundquatspinor}
\Psi\,=\,\left(\begin{array}{c}q_{1}\\q_{2}\end{array}\right)
\end{equation}
when using homogeneous coordinates $q_{1}$ and $q_{2}$ to relate the spinor 
$\Psi$ to points $q=q_{1}/q_{2}$ in the quaternionic projective `plane'. The 
appropriate matrix transformations can be identified with matrices ${\mathcal 
A}$ according to 
\begin{equation}
\label{eq:quatgruppe}
f(q)\,=\,\frac{a q+b}{c q+d}
\quad\longleftrightarrow\quad
{\mathcal A}\,=\,\left(\begin{array}{cc}a&b\\c&d\end{array}\right)
\,,\quad a,b,c,d\in\mbox{\bf H}\,,
\end{equation}
so that the composition of generalized M\"obius transformations 
(\ref{eq:quatmoeb}) is equivalent to simple matrix multiplication. The matrix 
groups in the quaternionic case $S^4$ $\to$ {\bf H} are the compact groups 
Sp(2) and its subgroup SU(2)$\times$SU(2) related to rotations of the sphere 
and the noncompact group SL(2,{\bf H}) describing general (noneuclidean) 
transformations.
\end{itemize}
However, we are faced with the problem that we don't have an appropriate 
quaternionic analysis yet. Thus, instead of investigating infinitesimal 
properties of the quaternionic transformations $f(q)$ on the basis of an 
analysis, we want to benefit from the theory of Lie groups which allows 
to represent and investigate local transformations as well. Due to the Lie 
algebra isomorphism sl(2,{\bf H}) $\cong$ su$*$(4) $\cong$ 
so(5,1)\index{sl(2,{\bf H})}, all infinitesimal (local) properties of $f(q)$ 
resp. SL(2,{\bf H}) may be equally well discussed in terms of so(5,1) on real 
representation spaces or in terms of su$*$(4) on complex representation spaces. 
This Lie algebra isomorphism is the basis of the algebraic hadron theory 
presented in section~\ref{sect4}.

Note, that the geometrical generalization given in this section is a direct 
mathematical generalization on the basis of the four division algebras {\bf R}, 
{\bf C}, {\bf H} and {\bf O}, denoting real numbers, complex numbers, 
quaternions and octonions, respectively. In this geometrical scheme, there is 
{\it no need} to refer to the physical motivation already given for SU$*$(4) 
in section~\ref{sect2} but all symmetry properties are derived from the 
projection onto quaternions.

\section{An algebraic theory}
\label{sect4}
% Modell mit Gruppen

If we use Lie theory to investigate infinitesimal and global properties of 
the quaternionic projective geometry, there exist two further possibilities 
to investigate quaternionic projective transformations (\ref{eq:quatgruppe}). 
The Lie algebra isomorphism sl(2,{\bf H}) $\cong$ su$*$(4) $\cong$ so(5,1) 
suggests investigations on real and complex representation spaces by means of 
the Lie algebras su$*$(4) and so(5,1). Taking all the relevant quaternionic, 
complex and real algebras and groups into account, an appropriate algebraic 
theory leads to the scheme given in figure~\ref{gruppenbild}. 

\begin{figure}[t]
\centerline{\epsfxsize=14cm \epsfbox{grupengl.eps}}
\caption{Hierarchy of algebras and groups relevant for low and intermediate 
energy physics of elementary particles.}
\label{gruppenbild}
\end{figure}

This algebraic theory comprises three possible reduction schemes related to 
the three division algebras {\bf H}, {\bf C} and {\bf R} involved in its 
representation theory. Using quaternions, the chain SL(2,{\bf H}) $\supset$ 
Sp(2) $\supset$ Sp(1)$\times$Sp(1) $\supset$ Sp(1) is possible where one should 
put special emphasis on the quaternionic projective space {\bf HP(1)} $\cong$ 
Sp(2)/Sp(1)$\times$Sp(1). The same symmetry transformations can be investigated 
in terms of an isomorphic complex representation theory by use of the chain 
SU$*$(4) $\supset$ USp(4) $\supset$ SU(2)$\times$SU(2) $\supset$ SU(2). Besides 
the possibility to discuss the projection USp(4)/SU(2)$\times$SU(2) in terms 
of complex numbers, we can use Weyl's unitary trick to relate the noncompact 
group SU$*$(4) to the compact group SU(4) which we used in section~\ref{sect2} 
as a nonrelativistic classification scheme of hadrons.

The reduction scheme on real spaces begins with the group SO(5,1) which is 
covered twice by SU$*$(4). In this chain, we have more possibilities to 
identify groups which are relevant in classical physics. If we look for the 
noncompact subgroups SO(4,1) and SO(3,1) of SO(5,1), these groups can be 
identified with the de Sitter and the Lorentz group. The de Sitter group allows 
an approach to the Poincar\'{e} and the Galilei group which is discussed in 
section~\ref{hinzupoinc}. The groups SO(6), SO(5) and SO(4) emerge by Weyl's 
unitary trick where SO(5) has been discussed in the context of linear 
relativistic wave equations (Bhabha equations), the Duffin-Kemmer-P\'{e}tiau 
ring \cite{fb}, \cite{kn}, and in Pauli's approach towards a unified theory 
\cite{pauli}. SO(4) is the symmetry group of meson transformations in the 
linear sigma model, whereas its covering group SU(2)$\times$SU(2) is used to 
define the appropriate fermion spinors or `nonlinear' meson representations. 
SO(6) has been used as a generalization of the SO(4) sigma model where the 
mesons have been associated with the regular representation \rhdmulti{15} of 
SO(6), and third rank spinor representations of the covering group SU(4) have 
been used to classify hadronic fermions \cite{daki2}.

Here, we cannot treat all the algebras and groups involved in the proposed 
scheme in detail. Instead, we focus on a more detailed discussion of the two 
Lie algebras su$*$(4) and so(5,1) on complex and real representation spaces 
which are equivalent to sl(2,{\bf H}) and thus describe the same infinitesimal 
`physics'. The Lie algebra su$*$(4) allows to define a Clifford product and 
can be related to the Dirac algebra $\Gamma$ on {\bf C}$_{4\times 4}$ whereas 
so(5,1) allows to obtain the Poincar\'{e} algebra. As further examples for the 
powerful quaternionic projective geometry, we discuss in section~\ref{chirdyn} 
the geometrically simple relations of the linear sigma model \cite{gmlevy} to 
typical nonlinear sigma models currently used in hadron physics and in 
section~\ref{classmatter} the classification according to the proposed theory.
Section~\ref{noncanondecomp} is devoted to the effective description of 
special relativity in terms of complex quaternions and their relation to 
the quaternionic projective theory as well as to the relation of space-time 
and internal degrees of freedom.

\subsection{Complex representation and su$*$(4)}
\label{diracalgebra}
% komplexe Darstellungstheorie & Clifford-Algebra

A general basis of {\bf C}$_{4\times 4}$ consists of 32 elements which we want 
to parametrize by tensor products of two quaternions $(q_{\alpha},q_{\beta})$ 
(which we denote by `qquaternions' \cite{dahmdiss}\index{qquaternions}) with 
complex coefficients. Using the embedding (\ref{eq:pauliquats}) of quaternions, 
we obtain a 16dimensional algebra where the first quaternion $q_{\alpha}$ 
determines the block structure and the second one, $q_{\beta}$, specifies the 
entry. Allowing further for qquaternions with complex coefficients, we obtain 
a 32dimensional algebra in {\bf C}$_{4\times 4}$ which is a complete matrix 
algebra.

However, heading towards a decomposition of this 32dimensional matrix algebra 
with respect to Lie group theory it is very useful to subdivide it using 
hermitean conjugation, i.e. according to the eigenspaces $+/-$ of the 
involution $^{+}$. In this context, it is of great use to define a new 
15dimensional {\it skew-hermitean} basis $\rhdya_{\alpha\beta}$, 
$\rhdya^{+}\,=\,-\rhdya$, $0\leq\alpha,\beta\leq 3$, in {\bf C}$_{4\times 4}$ 
by 
\begin{equation}
\label{eq:yakowdef}
\rhdya_{j0}\,=\,(q_{j},q_{0})\,,\quad
\rhdya_{0k}\,=\,(q_{0},q_{k})\,,\quad 
\rhdya_{jk}\,=\,i(q_{j},q_{k})\,,
\end{equation}
$1\leq j,k\leq 3$. The exponential mapping
\begin{equation}
\exp\,:\,\rhdya\,\longrightarrow\,\{U\}_{4\times 4}
\end{equation}
relates $\rhdya_{j0}$, $\rhdya_{0k}$ and $\rhdya_{jk}$ to unitary matrices $U$ 
in {\bf C}$_{4\times 4}$ and to the group SU(4).

In the complete 32dimensional matrix algebra {\bf C}$_{4\times 4}$, we can 
further identify the Lie algebra su$*$(4) according to the definition 
\cite{helgason1} 
\begin{equation}
\mbox{su$*$(4)}\,=\,\left\{
\left(
\begin{array}{cc}
A_{1} & A_{2}\\
A_{3} & A_{4}
\end{array}
\right)\,\right|\left.
\vphantom{\left(\begin{array}{c} A_{1}\\A_{3}\end{array}\right)}
\,A_{1},\ldots,A_{4}\in\mbox{\bf C}_{2\times 2}
\right\}
\end{equation}
if $A_{3}=-\overline{A}_{2}$, $A_{4}=\overline{A}_{1}$ and 
$\mbox{Tr}(A_{1}+\overline{A}_{1})=0$ where $\overline{A}_{i}$ denotes 
{\it complex} conjugation. In the skew-hermitean basis (\ref{eq:yakowdef}), 
we find the generators of SU$*$(4) given by
\begin{equation}
\label{eq:algebrasustern4}
\begin{array}{rcrrrrl}
su$*$(4) & = & \{\,\,\rhdya_{02}, & \rhdya_{10}, & \rhdya_{20}, 
& \rhdya_{30}, & \,\rhdya_{11},\\
& & \rhdya_{13}, & \rhdya_{21}, & \rhdya_{23}, & \rhdya_{31}, 
& \,\rhdya_{33}\\
& & \rhdya_{01}, & \rhdya_{03}, & \rhdya_{12}, & \rhdya_{22}, 
& \rhdya_{32}\,\,\}\,.
\end{array}
\end{equation}
Introducing two sets $G_{1}$ and $G_{2}$ by
\begin{equation}
\label{eq:geins}
\begin{array}{rcrrrrl}
G_{1} & = & \{\,\,\rhdya_{02}, & \rhdya_{10}, & \rhdya_{20},
& \rhdya_{30}, & \,\rhdya_{11},\\
& & \rhdya_{13}, & \rhdya_{21}, & \rhdya_{23}, & \rhdya_{31},
& \,\rhdya_{33}\,\,\}
\end{array}
\end{equation}
and
\begin{equation}
\label{eq:gzwei}
G_{2}\,=\,\{\,\,
\rhdya_{01},\,\rhdya_{03},\,\rhdya_{12},\,\rhdya_{22},\,\rhdya_{32}\,\,\}\,,
\end{equation}
the Lie algebras su$*$(4) and su(4) can be represented as
\begin{equation}
\label{eq:cartanzerl}
\begin{array}{lcl}
\mbox{su$*$(4)} & = & G_{1}\,\oplus\,iG_{2}\\
\mbox{su(4)} & = & G_{1}\,\oplus\,\hphantom{i}G_{2}
\end{array}
\end{equation}
which reflects the relation of su$*$(4) and su(4) by Weyl's unitary trick. The 
subalgebra $G_{1}$ is isomorphic to the Lie algebra sp(2,{\bf C}) and generates 
the maximal compact subgroup Sp(2) of SU$*$(4). The proof is straightforward by 
applying the definition of the Lie algebra sp(2,{\bf C}) \cite{helgason1},
\begin{equation}
\mbox{sp(2,{\bf C})}\,=\,\left\{\left(
\begin{array}{cc} A_{1} & A_{2}\\ A_{3} & A_{4}\end{array}
\right)\,\right|\left.
\vphantom{\left(\begin{array}{c} A_{1}\\A_{3}\end{array}\right)}
\,A_{1},\ldots,A_{4}\in\mbox{\bf C}_{2\times 2}\right\}\,,
\end{equation}
where $A^{T}_{2}=A_{2}$, $A^{T}_{3}=A_{3}$ and $A_{4}=-A^{T}_{1}$. Thus, with 
respect to hermitean conjugation we find the Cartan decomposition of su$*$(4)
generators according to 10 $\oplus$ 5 where the 10 skew-hermitean operators 
generate the compact subgroup Sp(2). Besides this decomposition, it is 
noteworthy to mention the nontrivial anticommutators of the operators 
$\rhdya_{\alpha\beta}$ in eq.~(\ref{eq:yakowdef}). If we calculate their 
commutation and anticommutation properties, the elements $\rhdya_{\alpha\beta}$ 
of the above Lie algebras fulfil the commutation relations
\begin{equation}
\begin{array}{lclclcl}
\left[\rhdya_{00},\rhdya_{\alpha\beta}\right] & = & 0,
& \quad &
\left[\rhdya_{0j},\rhdya_{0k}\right] & = & 2\,\epsilon_{jkl}\,\rhdya_{0l},\\

\left[\rhdya_{0j},\rhdya_{k0}\right] & = & 0,
& \quad &
\left[\rhdya_{0j},\rhdya_{kl}\right] & = & 2\,\epsilon_{jlm}\,\rhdya_{km},\\

\left[\rhdya_{j0},\rhdya_{k0}\right] & = & 2\,\epsilon_{jkl}\,\rhdya_{l0},
& \quad &
\left[\rhdya_{j0},\rhdya_{kl}\right] & = & 2\,\epsilon_{jkm}\,\rhdya_{ml},\\

\left[\rhdya_{jk},\rhdya_{lm}\right] & = & 
\multicolumn{5}{l}{2\,\delta_{jl}\epsilon_{kmn}\,\rhdya_{0n}\,
+\,2\,\delta_{km}\epsilon_{jln}\,\rhdya_{n0}\,.}
\end{array}
\end{equation}
where the unit matrix in {\bf C}$_{4\times 4}$ is denoted by $\rhdya_{00}$.
From these commutation relations it is obvious that the two three dimensional 
sets {$\rhdya_{0j}$} and {$\rhdya_{k0}$} generate an SU(2)$\times$SU(2) 
subgroup of SU(4). The anticommutation relations read as
\begin{equation}
\begin{array}{lclclcl}
\left\{\rhdya_{00},\rhdya_{\alpha\beta}\right\} & = & 
\hphantom{-}2\,\rhdya_{\alpha\beta},
& \quad &
\left\{\rhdya_{0j},\rhdya_{0k}\right\} & = & -2\,\delta_{jk}\,\rhdya_{00},\\

\left\{\rhdya_{0j},\rhdya_{k0}\right\} & = & -2i\,\rhdya_{kj},
& \quad &
\left\{\rhdya_{0j},\rhdya_{kl}\right\} & = & -2i\,\delta_{jl}\,\rhdya_{k0},\\

\left\{\rhdya_{j0},\rhdya_{k0}\right\} & = & -2\,\delta_{jk}\,\rhdya_{00},
& \quad &
\left\{\rhdya_{j0},\rhdya_{kl}\right\} & = & -2i\,\delta_{jk}\,\rhdya_{0l},\\

\left\{\rhdya_{jk},\rhdya_{lm}\right\} & = & 
\multicolumn{5}{l}{-2\,\delta_{jl}\delta_{km}\,\rhdya_{00}\,
+\,2i\,\epsilon_{jln}\epsilon_{kmp}\,\rhdya_{np}\,.}
\end{array}
\end{equation}

Furthermore, using the definition
\begin{equation}
\label{eq:cliffalg}
\gamma^{0}\,=\,i\rhdya_{03}\in i G_{2}\,,\quad
\gamma^{j}\,=\,\rhdya_{j1}\in G_{1}\,,
\end{equation}
the matrices $\gamma^{0},\gamma^{j}\in$ su$*$(4) fulfil the Clifford product
\begin{equation}
\frac{1}{2}\,\left\{\gamma^{0},\gamma^{0}\right\}\,=\,1_{4\times 4}\,,\quad
\frac{1}{2}\,\left\{\gamma^{0},\gamma^{j}\right\}\,=\,0\,,\quad
\frac{1}{2}\,\left\{\gamma^{j},\gamma^{k}\right\}\,=\,
-\delta_{jk}1_{4\times 4}\,,
\end{equation}
and the full Dirac algebra can be constructed according to 
\begin{equation}
\gamma_{5}\,=\,\rhdya_{02},\,\gamma_{5}\gamma^{0}\,=\,i\rhdya_{01},\,
\gamma_{5}\gamma^{j}\,=\,-\rhdya_{j3},\,\sigma^{0j}\,=\,i\rhdya_{j2},\,
\sigma^{jk}\,=\,\epsilon_{jkl}\rhdya_{l0}\,.
\end{equation}
Thus, adding the unit element $1_{4\times 4}$ to the adjoint representation 
of su$*$(4), the algebra is isomorphic to the Dirac algebra, and the sixteen 
coefficients (`fields') of the decomposition
\begin{equation}
\label{eq:diracdecomp}
\Gamma\,=\,s 1_{4\times 4}\,+\,p\gamma_{5}\,+\,v_{\mu}\gamma^{\mu}\,+\,
a_{\mu}\gamma_{5}\gamma^{\mu}\,+\,f_{\mu\nu}\sigma^{\mu\nu}
\end{equation}
can be associated with {\it real} coefficients of the regular representation of 
the group SU$*$(4). Besides the nice possibility to identify various `fields' 
used in perturbative and nonperturbative/effective models, this theory allows 
to determine the appropriate global {\it and} infinitesimal transformation 
properties {\it exactly}. Furthermore, the application of Lie group theory 
allows to handle {\it finite} transformations as well as infinitesimal (local) 
transformations which is interesting with respect to summing up a complete 
perturbation series.

\subsection{Real representation and so(5,1)}
\label{hinzupoinc}
% hin zur Poincar\'e-Algebra

The relation of quaternionic projective theory to observables in classical 
physics can be established by representing the isomorphic Lie algebra so(5,1) 
on real spaces. 

The relevant algebra of classical dynamics is the Poincar\'{e} algebra 
${\mathcal P}$ which generates finite (macroscopic) space-time transformations. 
However, it is wellknown \cite{levynahas} that only two Lie algebras can be 
contracted\index{contraction} \cite{gilmore} to the Poincar\'{e} algebra, 
namely the de Sitter algebra so(4,1) and the anti-de Sitter algebra so(3,2). 
Thus, as a direct possibility, we can identify the subalgebra so(4,1) of 
so(5,1) straightforward with the de Sitter algebra to obtain the Poincar\'{e} 
algebra by contraction, i.e. in the limit of vanishing curvature. This 
interpretation allows to identify the ten generators of Poincar\'{e} space-time 
transformations in a contracted (projective) limit of quaternionic generators 
in sl(2,{\bf H}). In this sense, classical dynamics on real representation 
spaces and Dirac theory on complex representation spaces can be understood as 
{\it two facets} of one and the same quaternionic projective theory. It is 
noteworthy, that the Dirac algebra is {\it isomorphic} to the quaternionic 
projective theory in terms of su$*$(4) whereas classical dynamics, described 
by means of the Poincar\'{e} algebra, appears after an additional contraction, 
i.e. in a special (projective) limit. Because su$*$(4) (resp. su(4) as given in 
section~\ref{sect2}) already describes internal symmetry (flavour) degrees of 
freedom ${\mathcal I}$, quaternionic theory suggests that one {\it shouldn't} 
handle dynamic and internal symmetry by simply assuming direct/semidirect 
products like ${\mathcal P}\times {\mathcal I}$. Moreover, space-time and 
internal symmetry transformations are connected by commutation/anticommutation 
relations of the generators of sl(2,{\bf H}) $\cong$ su$*$(4) $\cong$ so(5,1).

In real representation theory, we can use two mathematical approaches. Either 
the generators are represented by matrices acting on appropriate real vector 
spaces or we can choose representations of the generators in terms of 
differential operators acting on spaces of (square-integrable) functions. The 
second approach allows to discuss transformation laws in terms of differential 
equations and appropriate polynomial systems as solutions of these equations.
Thus, if we represent the generators of the Lie algebra so(5,1) and their 
subalgebras as differential operators \cite{helgason1}, \cite{gilmore}, we 
can introduce appropriate coordinate systems and relate the differential 
equations to dynamic laws known from classical physics.

\subsection{Chiral Dynamics revisited}
\label{chirdyn}
After the identification of the SU(2)$\times$SU(2) meson representation 
(\ref{eq:zentral}) as unit quaternion, there exists an elegant geometrical 
approach towards linear and nonlinear sigma models and their relations. 
Restricting the quaternions in eq.~(\ref{eq:quatmoeb}) to a special 
structure, $b=c=0$, $a,d$ $\in$ U(1,{\bf H}) and $q=U$ $\in$ U(1,{\bf H}), 
the generalized M\"obius transformation takes the form
\begin{equation}
\label{eq:abgeschlossen}
f(U)\,=\,a U d^{-1}\,.
\end{equation}
The treatment of quaternions differs from the complex case in that the 
quaternions $U$ and $d^{-1}$ ($=d^{+}$ for $d$ $\in$ U(1,{\bf H})) in general 
do not commute. Therefore, the appropriate symmetry group in the case of the 
quaternionic circle is U(1,{\bf H})$\times$U(1,{\bf H}) respectively 
SU(2)$\times$SU(2) if we use the isomorphism SU(2) $\cong$ U(1,{\bf H}). Thus, 
invariance under the symmetry group SU(2)$\times$SU(2) reflects on complex 
representation spaces the fact that U(1,{\bf H}) is closed under quaternionic 
multiplication, $f(q)$ $\in$ U(1,{\bf H}) in eq.~(\ref{eq:abgeschlossen}), 
and the group transformations map unit quaternions onto unit quaternions (they 
act on the quaternionic `unit circle'). 

The product structure of the `chiral' group stems from the quaternionic 
projective transformation law (\ref{eq:quatgruppe}) if we use the description 
of quaternions in the projective plane and from the fact that quaternions in 
general do not commute. The decoupling of the two SU(2) groups in the `chiral' 
group becomes even more obvious in terms of the matrix representation 
(\ref{eq:quatgruppe}) if we choose as above $b=c=0$ and $a,d$ $\in$ 
SU(2,{\bf C}). Note, that in this theory there is {\it no} necessity to 
introduce `handedness' with respect to a chiral structure of objects or to 
think about mass of fermions and bosons. The `chiral' structure originates 
directly from the higher hypercomplex and noncommutative geometry. In the 
complex case, this `chiral' structure doesn't exist because the symmetry group 
in the projective plane reduces to U(1,{\bf C}) due to the commutativity of 
complex numbers. Thus, we are left with the U(1,{\bf C}) symmetry which allows 
to choose a complex phase like in quantum mechanics\index{chiral product 
structure}.

Geometrically, U(1,{\bf H})$\times$U(1,{\bf H}) symmetry stems from restricted 
rotations of the sphere $S^{4}$ with respect to a fixed axis through its north 
pole, i.e. if the symmetry group SO(5) (resp. its covering Sp(2)) is restricted 
to SO(4) (resp. its covering SU(2)$\times$SU(2)). This geometry justifies the 
identification
\begin{equation}
\label{eq:expsummen}
\sigma\,=\,\cos\varphi\,,\quad\vec{\pi}\,=\,\hat{\varphi}\sin\varphi
\end{equation}
already given in (\ref{eq:zentral}) in the context of linear and nonlinear 
representations of unit quaternions. 

As discussed at the beginning of this section, the norm 
(\ref{eq:quaternionnorm}),
\[
||U||^{2}\,=\,\frac{1}{2}\,\mbox{\bf Tr}\left(U^{+}U\right)\,=\,
\sigma^{2}+\vec{\pi}^{\,2}\,=\,1_{2\times 2}
\]
of a unit quaternion $U$ is conserved under U(1,{\bf H})$\times$U(1,{\bf H}). 
The invariance of this norm, however, can be equally well interpreted in terms 
of SO(4) acting on {\bf R}$^{4}$ (respectively the sphere $S^{3}$) by 
introducing a four dimensional real vector $B=\left(\sigma,\vec{\pi}\right)$
such that
\begin{equation}
||U||^{2}\,=\,\frac{1}{2}\,\mbox{\bf Tr}\left(U^{+}U\right)\,=\,B^{2}\,.
\end{equation}
Thus, the linear sigma model \cite{gmlevy} covers a special aspect of 
quaternionic projective theory, and the three parameters $\varphi_{j}$ of the 
nonlinear representation $U$ in eq.~(\ref{eq:zentral}) can be related to an 
appropriate linear parametrization by using coordinates in {\bf R}$^{4}$ which 
are related to angles of $S^{3}$ (see for example the set of coordinates in 
\cite{boyer}).

However, it is obvious that expansions of $U$ in terms of parameters $\varphi$ 
as used in nonlinear effective hadron models \cite{leutwyler1} spoil this 
geometrical concept. Since the transcendental functions in (\ref{eq:expsummen}) 
respectively the `fields' $\sigma$ and $\vec{\pi}$ can be interpreted as 
{\it complete} sums of an even and an odd power series in $\varphi$, the 
linear sigma model yields a {\it complete} description of {\it mesonic} 
properties as well as a nonlinear theory in terms of $U$. This is {\it not} 
true for any expansions of $U$ in terms of $\varphi_{j}$ up to a certain {\it 
finite} order. The description of fermions with respect to general SU$*$(4) or 
Sp(2) transformations has to use quaternions, or, in an appropriate symmetry 
reduction caused by a fixed rotation axis of the sphere $S^{4}$, fermion 
spinors can be described by representations of the group SU(2)$\times$SU(2) 
like in Chiral Dynamics. Moreover, (\ref{eq:expsummen}) allows for a direct 
identification of flavour SU(2) quantum numbers in the reduction scheme 
SU(2)$\times$SU(2)/SU(2). Since the `chiral' product structure of 
SU(2)$\times$SU(2) originates from the noncommutative character of 
quaternionic projective transformations (\ref{eq:quatmoeb}) there is no 
further geometrical meaning in a naive generalization to arbitrary 
SU($n$)$\times$SU($n$) groups for $n>2$.

\subsection{Representation theory and the classification of matter fields}
\label{classmatter}
The relation of the Dirac algebra $\Gamma$ and the Lie group SU$*$(4) in 
section~\ref{diracalgebra} suggests to identify massive matter fields with 
appropriate representations of SU$*$(4). 

Due to the description (\ref{eq:fundquatspinor}) of the fundamental Dirac 
spinor in terms of quaternions, quantum electrodynamics suggests to associate 
the electron with the fundamental representation of SU$*$(4). The quaternionic 
projective approach then allows investigations of gauge theories as well as 
investigation of classical theories by appropriate identifications of the 
potentials respectively the classical fields $\vec{E}$ and $\vec{B}$ in 
(\ref{eq:diracdecomp}). Thus, tracing the coefficients in 
(\ref{eq:diracdecomp}) back to the generalized M\"obius transformation 
(\ref{eq:quatmoeb}) one is lead to a geometrical interpretation of QED in 
terms of quaternionic projective transformations.

As motivated in sections~\ref{dynamclass} and \ref{su4class}, higher spinorial 
representations of SU$*$(4) should be associated with massive hadrons in the 
particle spectrum. Appropriately, the {\it third} rank symmetric spinor 
$\Psi^{\alpha\beta\gamma}$ describes the 20 nucleon/Delta fermions whereas 
the regular representation describes the massive meson fields $\vec{\pi}$, 
$\omega$ and $\vec{\rho}$. Due to this classification scheme of the matter 
fields, there is no need to use the concept of a spontaneously broken 
SU(2)$\times$SU(2) symmetry with necessarily massless Goldstone pions and an 
additional explicit chiral symmetry breaking to restore the pion mass in order 
to explain the pseudovector coupling of pions to fermions. Geometrically, it 
is interesting that the fundamental spinor corresponds to a point in the 
quaternionic projective plane whereas the higher spinorial representations 
correspond to extended objects with certain symmetry properties and thus 
have additional degrees of freedom.

\subsection{A noncanonical decomposition of SU$*$(4) and space-time}
\label{noncanondecomp}
As a further interesting feature of SU$*$(4) representation theory, we want 
to connect quaternionic projective theory to the wellknown description of 
space-time and relativistic dynamics in terms of quaternions with complex 
coefficients \cite{blaton}, \cite{blaschke}, so called biquaternions 
\cite{clifford}\index{biquaternions}.

Therefore, we map su$*$(4) {\it not} by the canonical mapping 
$\exp:G_{1}\oplus i G_{2}\to\mbox{SU$*$(4)}$ onto the group but we focus on 
the properties of 
\begin{equation}
\label{eq:spacetime}
\underline{X}\,=\,
\exp\left(
-\alpha_{1}i\rhdya_{12}-\alpha_{2}i\rhdya_{22}-\alpha_{3}i\rhdya_{32}
\right)
\end{equation}
emerging in the noncanonical parametrization
\begin{equation}
\label{eq:noncanonmap}
\begin{array}{rcl}
g\in\mbox{SU$*$(4)} & = & 
\exp\left(\beta_{1}i\rhdya_{01}\right)\,
\exp\left(\beta_{2}i\rhdya_{03}\right)\,
\exp\left(\left\{G_{1}\right\}\right)\\
& & *\,\exp\left(
-\alpha_{1}i\rhdya_{12}-\alpha_{2}i\rhdya_{22}-\alpha_{3}i\rhdya_{32}
\right)\,.
\end{array}
\end{equation}
The elements $P\in\left\{i\rhdya_{01},i\rhdya_{03}\right\}$ parametrized by 
real coefficients $\beta_{1}$ and $\beta_{2}$ fulfil $P^{2}=1$ and thus give 
rise to four projection operators. The (symbolic) exponential on the rhs of 
the first equation in (\ref{eq:noncanonmap}) maps the Lie algebra sp(2,{\bf C}) 
onto the maximal compact subgroup Sp(2) of SU$*$(4). The additional negative 
signs of the real parameters $\alpha_{j}$ are introduced in the argument of 
the exponential (\ref{eq:spacetime}) to simplify calculations when using the 
qquaternionic basis defined in section~\ref{diracalgebra} instead of the 
skew-hermitean basis (\ref{eq:yakowdef}).

Rewriting (\ref{eq:spacetime}) in terms of qquaternions, we find
\begin{equation}
\label{eq:finitespacetime}
\begin{array}{rcl}
\underline{X} & = & 
\exp\left(
\alpha_{1}(q_{1},q_{2})+\alpha_{2}(q_{2},q_{2})+\alpha_{3}(q_{3},q_{2})
\right)\\
& \to & t (q_{0},q_{0})\,+\,x_{1}(q_{1},q_{2})\,+\,
x_{2}(q_{2},q_{2})\,+\,x_{3}(q_{3},q_{2})\,.
\end{array}
\end{equation}
Furthermore, $(q_{0},q_{0})$ commutes with the $(q_{j},q_{2})$ and 
$(q_{j},q_{2})$ is isomorphic to $(q_{2},q_{j})$ which can be seen by 
exchanging the order of indices and using the symmetry of qquaternionic 
multiplication. Therefore, we can associate the basis elements $(q_{2},q_{j})$ 
related to the {\it finite} coordinates $x_{j}$ with the matrix representations
\begin{equation}
(q_{2},q_{j})\,=\,\left(
\begin{array}{cc}
0 & -q_{j}\\
q_{j} & 0
\end{array}
\right)
\end{equation}
due to the construction scheme of qquaternions. Using the $2\times 2$ matrix 
representation 
\begin{equation}
\mbox{\bf 1}\,=\,\left(\begin{array}{cc} 1 & 0\\ 0 & 1\end{array}\right)\,,
\quad i\,=\,\left(\begin{array}{cc} 0 & -1\\ 1 & 0\end{array}\right)\,,
\quad i^{2}\,=\,-\,\mbox{\bf 1}
\end{equation}
of the imaginary (commutative) unit $i$, it is obvious that the event 
$\underline{X}$ in (\ref{eq:spacetime}) emerging in the noncanonical 
decomposition (\ref{eq:noncanonmap}) of SU$*$(4) group elements is isomorphic 
to the wellknown description of (hermitean) space-time in terms of 
biquaternions,
\begin{equation}
\label{eq:biquats}
\underline{X}\,=\,x_{0}q_{0}\,+\,iq_{j}\,x_{j}\,.
\end{equation}
Thus, it is possible to express all transformations of the Lorentz group, i.e. 
rotations and boosts, in a common framework \cite{dahmdiss} on the basis of 
noncanonical decompositions of qquaternionic mappings. 

Furthermore, because the `norm' 
$||\underline{X}||^{2}=x^{2}_{0}-x^{2}_{1}-x^{2}_{2}-x^{2}_{3}$
of space-time events as given in (\ref{eq:finitespacetime}) resp. 
(\ref{eq:biquats}) is conserved under Lorentz transformations, the Wick 
rotation as a complexification of the time component allows to factor out and 
omit the overall sign, 
\begin{equation}
\label{eq:compactnorm}
||\underline{X}||^{2}\,\to\,
\left(ix_{0}\right)^{2}-x^{2}_{1}-x^{2}_{2}-x^{2}_{3}\,=\,
-\left(x^{2}_{0}+x^{2}_{1}+x^{2}_{2}+x^{2}_{3}\right)\,.
\end{equation}
Appropriately, the norm on the rhs of (\ref{eq:compactnorm}) is conserved under 
SO(4) transformations which reflects the relation of the Lorentz group SO(3,1) 
and the compact group SO(4) by Weyl's unitary trick. Thus, we can express 
(Wick rotated) space-time events in terms of four dimensional vectors with 
real coordinates which are related phenomenologically to the division algebra 
of quaternions, to the symmetry group SO(4) and to Gegenbauer polynomials. 

This approach has the further nice feature that it allows to explain some 
properties of quantum mechanics and the first quantization \cite{dahmdiss} 
although we cannot explain Planck's constant with respect to the simple, 
restricted algebraical considerations given above. Applying a Wick rotation to 
space-time, the conservation of the norm (\ref{eq:compactnorm}) of `space-time' 
events leads to the symmetry group SO(4) acting on a four dimensional real 
euclidean space. However, fixing the time component in the representation 
(\ref{eq:finitespacetime}), i.e. looking only for stationary problems at fixed 
time, one benefits from the direct product decomposition of SO(4) according to
SO(4) $\cong $ SO(3)$\times$SO(3). Instead of SO(4) transformations mixing 
(Wick-rotated) time with the three space components, the remaining 
transformations in coordinate space at fixed time respect automatically SO(3) 
rotation symmetry. But this is exactly the point where nonrelativistic quantum 
mechanics starts by solving differential equations for square-integrable 
functions on {\bf R}$^{3}$.

\section{Summary and Outlook}
\label{sect6}

We have presented a generalization of `effective' hadron models towards an 
algebraic theory based on the division algebra of quaternions. This theory 
allows from `first principles' to embed Chiral Dynamics completely into 
quaternionic transformations, to identify particle representations and quantum 
numbers and to relate various effective hadron models on a geometrical basis. 
Furthermore, real and complex representations of quaternionic projective 
theory allow to identify symmetry transformations in classical physics as well 
as transformations of quantum field theory as two facets of one and the same 
quaternionic theory (section~\ref{sect4}). In quaternionic theory, space-time 
events and internal degrees of freedom are treated in an unified framework of 
quaternionic transformations like in other approaches based on Clifford 
algebras \cite{keller}.

Besides the topics covered in section~\ref{sect4}, there is a large variety 
of other very interesting features comprised in the algebraic theory presented 
in figure~\ref{gruppenbild}. Especially with respect to the mass spectrum and 
a discussion of relativistic wave equations in terms of induced representations 
\cite{niedoraif}, the quaternionic projective theory allows to focus on coset 
decomposition like SU$*$(4)/USp(4) or the projective space {\bf HP(1)}. For the 
framework of Chiral Dynamics is completely embedded in this algebraic theory 
on the basis of quaternionic transformations and their representations, it 
seems possible to handle relativistic transformations of physical hadrons in a 
complete and exact framework. In this context, explicit calculations of cross 
sections and further observables can be achieved by using representation 
theory of the groups SU$*$(4) and SO(5,1) so that local as well as global 
transformation properties of the representations are exactly determined. 

Comparison of these calculations with experiments should be done in an energy 
range where it is possible to discriminate between the compact group SU(4) and 
the related noncompact group SU$*$(4) but without having too much influence of 
higher resonances. Practically, this restricts investigations to the dynamics 
of the $\pi N\Delta$-system where the quaternionic projective theory has its 
roots. However, in this energy regime we have access to experimental data 
by high precision experiments which are possible at the electron accelerator 
MAMI \cite{walcher2}.

With respect to an algebraic construction {\bf R} $\to$ {\bf C} $\to$ {\bf H} 
$\to$ {\bf O} of the division algebras, {\it one} further extension of the 
above quaternionic geometry is possible on the basis of octonions. However, 
due to the nonassociative structure of the algebra of octonions \cite{dixon} 
such a model cannot be discussed completely in terms of groups and by simple 
matrix representations but one has to subdivide octonions to cover certain 
aspects with tools like group theory or standard representation theory.

\acknowledgements
The author wants to thank M. Kirchbach for many helpful discussions during 
the investigations related to this work, J. Keller for the invitation to 
participate in the conference and for the enormous hospitality in M\'exico, 
Z. Oziewicz for the possibility to present the investigations during the 
conference and S. Rodr\'\i{}guez-Romo for the hospitality at the `Center of 
Theoretical Research', Cuautitl\'an, M\'exico.


\begin{thebibliography}{}
\bibitem[\protect\citeauthoryear{Adler and Dashen}{1968}]{adldash} 
Adler S. L. and R. F. Dashen [1968], ``Current Algebras'', 
W. A. Benjamin, Inc., New York
\bibitem[\protect\citeauthoryear{Blaschke}{1959}]{blaschke} 
Blaschke W. [1959], Anwendung dualer Quaternionen auf Kinematik, 
{\it Ann. Acad. Scient. Fennicae} {\bf A I 250}, 3
\bibitem[\protect\citeauthoryear{Blaton}{1935}]{blaton} 
Blaton J. [1935], Quaternionen, Semivektoren und Spinoren,
{\it Zeitschr. f. Physik} {\bf 95}, 337
\bibitem[\protect\citeauthoryear{Boyer}{1971}]{boyer} 
Boyer C. P. [1971], Matrix Elements of the Most Degenerate Continuous Principal 
Series of Representations of SO(p,1), {\it J. Math. Phys.} {\bf 12}, 1599
\bibitem[\protect\citeauthoryear{Callan {\it et al.}}{1969}]{wesszum2} 
Callan C. G., S. Coleman, J. Wess and B. Zumino [1969], Structure of 
Phenomenological Lagrangians. II., {\it Phys. Rev.} {\bf 177}, 2247
\bibitem[\protect\citeauthoryear{Clifford}{1878}]{clifford} 
Clifford W. K. [1878], Applications of Grassmann's Extensive Algebra, 
{\it Amer. J. Math.} {\bf 1}, 350
\bibitem[\protect\citeauthoryear{Coleman {\it et al.}}{1969}]{wesszum1} 
Coleman S., J. Wess and B. Zumino [1969], Structure of Phenomenological 
Lagrangians. I., {\it Phys. Rev.} {\bf 177}, 2239
\bibitem[\protect\citeauthoryear{Dahm {\it et al.}}{1994}]{daki1} 
Dahm R., M. Kirchbach and D. O. Riska [1994], Wigner Supermultiplets in 
$\pi^{0}$-Photo\-pro\-duc\-tion at Threshold, Proceed. ICSMP 1993, Dubna; 
{\it Int. J. Mod. Phys.} {\bf B10} (1995) (in press)
\bibitem[\protect\citeauthoryear{Dahm}{1994}]{dahmdiss} 
Dahm R. [1994], ``Spin-Flavour-Symmetrien und das $\pi N\Delta$-System'', 
PhD thesis, KPH 20/94, Mainz (to be published)
\bibitem[\protect\citeauthoryear{Dahm and Kirchbach}{1995}]{daki2} 
Dahm R. and M. Kirchbach [1995], Linear Wave Equations and Effective 
Lagrangians for Wigner Supermultiplets, {\it Int. J. Mod. Phys.} {\bf A10}, 
4225
\bibitem[\protect\citeauthoryear{Dixon}{1994}]{dixon} 
Dixon G. M. [1994], ``Division Algebras: Octonions, Quaternions, Complex 
Numbers and the Algebraic Design of Physics'', Kluwer Academic Publishers, 
Dordrecht
\bibitem[\protect\citeauthoryear{Fabri {\it et al.}}{1967}]{fabpic} 
Fabri E., L. E. Picasso and F. Strocchi [1967], Quantum Field Theory and 
Approximate Symmetries, {\it Nuovo Cimento} {\bf 48A}, 376 
\bibitem[\protect\citeauthoryear{Fischbach {\it et al.}}{1974}]{fb} 
Fischbach E., J. D. Louck, M. M. Nieto and C. K. Scott [1974], The Lie Algebra 
so(N) and the Duffin--Kemmer--Petiau Ring, {\it J. Math. Phys.} {\bf 15}, 60
\bibitem[\protect\citeauthoryear{Franzini and Radicati}{1963}]{franz1} 
Franzini P. and L. A. Radicati [1963], On the Validity of Supermultiplet Model, 
{\it Phys. Lett.} {\bf 6}, 322
\bibitem[\protect\citeauthoryear{Gel'fand {\it et al.}}{1963}]{gms} 
Gel'fand I. M., R. A. Minlos and Z. Ya. Shapiro [1963], ``Representations of 
the Rotation and Lorentz Groups and Their Application'', Pergamon Press, Oxford
\bibitem[\protect\citeauthoryear{Gell-Mann and Levy}{1960}]{gmlevy} 
Gell-Mann M. and M. Levy [1960], Axial Vector Current in Beta Decay, 
{\it Nuovo Cimento} {\bf 16}, 705 
\bibitem[\protect\citeauthoryear{Gilmore}{1974}]{gilmore} 
Gilmore R. [1974], ``Lie Groups, Lie Algebras and Some of Their Applications'', 
John Wiley \& Sons, New York
\bibitem[\protect\citeauthoryear{Haag and Kastler}{1964}]{haka} 
Haag R. and D. Kastler [1964], An Algebraic Approach to Quantum Field Theory, 
{\it J. Math. Phys.} {\bf 5}, 848
\bibitem[\protect\citeauthoryear{Helgason}{1962}]{helgason1} 
Helgason S. [1962], ``Differential Geometry and Symmetric Spaces'', 
Academic Press, New York
\bibitem[\protect\citeauthoryear{Joos and Weimar}{1976}]{joosweim} 
Joos H. and E. Weimar [1976], On the Covariant Description of Spontaneously 
Broken Symmetry in General Field Theory, {\it Nuovo Cimento} {\bf 32A}, 283
\bibitem[\protect\citeauthoryear{Keller}{1995}]{keller} 
Keller J. [1995], private communications, see also the contribution at this 
conference
\bibitem[\protect\citeauthoryear{Kirchbach and Riska}{1991}]{kiri} 
Kirchbach M. and D. O. Riska [1991], On the Consistency of Effective Models of 
Nucleon Structure with Chiral Symmetry, {\it Nuovo Cimento} {\bf 104}, 1837 
\bibitem[\protect\citeauthoryear{Krajcik and Nieto}{1977}]{kn} 
Krajcik R. A. and M. M. Nieto [1977], Bhabha First-Order Wave 
Equations. VII. Summary and Conclusions, {\it Phys. Rev.} {\bf D15}, 445
\bibitem[\protect\citeauthoryear{Leutwyler}{1991}]{leutwyler1} 
Leutwyler, H. [1991], in: H. Mitter, H. Gausterer (eds.) ``Recent Aspects of 
Quantum Fields'', Springer, Berlin (Lecture Notes in Physics, Vol. 396) 
\bibitem[\protect\citeauthoryear{Levy-Nahas}{1967}]{levynahas} 
Levy-Nahas M. [1967], Deformation and Contraction of Lie Algebras, 
{\it J. Math. Phys.} {\bf 8}, 1211
\bibitem[\protect\citeauthoryear{Niederer and O'Raifeartaigh}{1974}]{niedoraif} 
Niederer U. H. and L. O'Raifeartaigh [1974], Realizations of the Unitary 
Representations of the Inhomogeneous Space-Time Groups, {\it Fortschr. Phys.} 
{\bf 22}, 111; {\it Fortschr. Phys.} {\bf 22}, 131
\bibitem[\protect\citeauthoryear{Oehme}{1965}]{oehme} 
Oehme R. [1965], Current Algebras and Approximate Symmetries, 
{\it Phys. Rev.} {\bf 143}, 1138
\bibitem[\protect\citeauthoryear{Pais}{1964}]{pais1} 
Pais A. [1964], Implications of Spin--Unitary Spin Independence, 
{\it Phys. Rev. Lett.} {\bf 13}, 175
\bibitem[\protect\citeauthoryear{Pauli}{1933}]{pauli} 
Pauli, W. [1933], \"Uber die Formulierung der Naturgesetze mit f\"unf 
homogenen Koordinaten, {\it Ann. Phys. (Lpzg.)} {\bf 18}, 305; {\it Ann. Phys. 
(Lpzg.)} {\bf 18}, 337
\bibitem[\protect\citeauthoryear{Penrose and Rindler}{1990}]{penrind} 
Penrose R. and W. Rindler [1990], ``Spinors and Space-Time'', 
Cambridge (Cambridge Monographs on Mathematical Physics)
\bibitem[\protect\citeauthoryear{Sudarshan}{1965}]{sudarsh} 
Sudarshan E. C. G. [1965], Theory of Approximate Symmetries, 
in: Salam A. (Dir.), ``High-Energy Physics and Elementary Particles'', Vienna
\bibitem[\protect\citeauthoryear{Walcher}{1994}]{walcher2} 
Walcher T. [1994], The Mainz Microtron MAMI: A Facility portrait with a 
glimpse at first results, {\it Nucl. Phys. News} {\bf 4}, No. 2, 5
\bibitem[\protect\citeauthoryear{Weinberg}{1968}]{weinberg1} 
Weinberg S. [1968], Nonlinear Realizations of Chiral Symmetry, 
{\it Phys. Rev.} {\bf 166}, 1508
\bibitem[\protect\citeauthoryear{Wigner}{1937}]{wigner1} 
Wigner E. P. [1937], On the Consequences of the Symmetry of the Nuclear 
Hamiltonian on the Spectroscopy of Nuclei, {\it Phys. Rev.} {\bf 51}, 106
\end{thebibliography}
\end {document}